\documentclass[aps,prd,twocolumn,floatfix,nofootinbib,showpacs,superscriptaddress]{revtex4}
\usepackage{graphicx,amsmath,amssymb,bm}
\usepackage{color}
\usepackage[utf8]{inputenc}
\definecolor{purple}{rgb}{0.5,0,0.5}
\definecolor{blue}{rgb}{0.0,0,0.9}
\usepackage[colorlinks=true, pdfstartview=FitV, citecolor= purple, linkcolor = blue,urlcolor=blue]{hyperref}

\begin{document}

\title{Mass spectrum and decay constants of radially excited vector mesons}
\author{Fredy F. Mojica}
\email{fremo22@gmail.com}
\author{Carlos E. Vera}
\email{cvera@ut.edu.co}
\affiliation{Departamento de F\'isica, Facultad de Ciencias, Universidad del Tolima, 730006299, Ibagu\'e, Colombia}
\author{Eduardo~Rojas}
\email{rojas@gfif.udea.edu.co}
\affiliation{Instituto de F\'isica, Universidad de Antioquia, Calle 70 No.~52-21, Medell\'in, Colombia}
\author{Bruno~El-Bennich}
\email{bruno.bennich@cruzeirodosul.edu.br}
\affiliation{Laborat\'orio de F\'isica Te\'orica e Computacional, Universidade Cruzeiro do Sul, Rua Galv\~ao Bueno, 868, 01506-000 S\~ao Paulo, SP, Brazil}

\begin{abstract}

We calculate the masses and weak decay constants of flavorless  ground and radially excited $J^P=1^-$ mesons and the corresponding quantities for the $K^*$,  within a Poincar\'e covariant continuum 
framework based on the Bethe-Salpeter equation. We use in both, the quark's gap equation and the meson bound-state equation, an infrared massive and finite interaction
in the leading symmetry-preserving truncation. While our numerical results are in rather good agreement with experimental values where they are available, no single 
parametrization of the QCD inspired interaction reproduces simultaneously the ground and excited mass spectrum, which confirms earlier work on pseudoscalar mesons. 
This feature being a consequence of the lowest truncation, we pin down the range and strength of the interaction in both cases to identify common qualitative features 
that may help to tune future interaction models beyond the rainbow-ladder approximation.

\end{abstract}

\maketitle

\section{Introduction  \label{intro}}

Vector mesons play an important role in the physics of strong interactions and hadron phenomenology. Since these mesons and the photon share the
same quantum numbers, $J^{PC}=1^{--}$ , flavorless neutral vector mesons can directly couple to the photon via an electromagnetic current. Historically, 
this led to the {\em vector meson dominance model\/}. Compared to other mesons their production can be measured with very high precision, for instance
in electron-positron collisions via the process $e^+e^- \to \gamma^* \to \bar qq$ which provides a much cleaner signal than hadronic reactions. Notwithstanding 
complications with hadronic final states, vector mesons are abundant decay products in electroproduction of excited nucleons~\cite{Aznauryan:2012ba,Cloet:2013jya}, 
$N^*$, and exclusive vector meson production reactions are responsible for a large fraction of the total hadronic cross section and precede di-lepton decays 
in relativistic heavy-ion collisions~\cite{Goncalves:2016sqy}. In flavor physics, exclusive $B$ decays with final-state vector mesons, e.g. $B\to V\pi$, $B \to V\ell \nu_\ell$, 
$B\to V\mu^+\mu^-$ or  $B \to V\gamma$, are a central component of the LHC$b$ experimental program and the  $B \to K^*\mu^+\mu^-$ and 
$B_s \to K^*\mu^+\mu^-$ decays are of particular interest, as their angular distributions are very sensitive probes of new physics~~\cite{ElBennich:2006yi,ElBennich:2009da,
Leitner:2010fq,ElBennich:2011gm,ElBennich:2012tp,Paracha:2014wra}. In the latter cases, a precise knowledge of the pseudoscalar and vector meson light-cone distribution 
amplitudes is essential~\cite{Braun:2016wnx}. 

A characteristic feature of the $1^{--}$ ground-state vector mesons is their predominant occurrence as pure $\bar qq$ states: in the case of the 
$\omega(782)$ and $\phi(1020)$ mesons the vector flavor-nonet mixing angle is close to ideal mixing, i.e. $\phi(1020)$  is nearly a pure $|\bar ss\rangle$ 
state and $\omega =(\bar uu  +\bar dd)/\surd{2}$. This ideal mixing is not evident in pseudoscalar and scalar meson multiplets. Thus, vector mesons as decay 
products of heavier non-vector mesons are a very good probe of their flavor content measured in their respective decay rates into different types of  mesons. 

The study of vector mesons is complementary to that of light pseudoscalar mesons, the Goldstone bosons, as their masses are more in agreement with the sum of 
their typical constituent quark masses. This stands in contrast to the light pseudoscalar's properties best described by the dichotomy of dynamical chiral symmetry 
breaking (DCSB), which generates a light-quark mass consistent with typical empirical constituent masses~\cite{ElBennich:2009vx,ElBennich:2012ij,daSilva:2012gf,
deMelo:2014gea,Yabusaki:2015dca} even in the chiral limit,  yet also produces a very light Goldstone boson due to explicit breaking of chiral symmetry of non-zero 
current-quark masses. These characteristic features  of the light pseudoscalar octet are dictated by an axialvector Ward-Green-Takahashi identity which relates 
dynamical quantities in the chiral limit. Most chiefly, it implies that the leading Lorentz covariant in the pseudoscalar quark-antiquark $\gamma_5$ channel is equal 
to $B(p^2)/f_\pi$, where $B(p^2)$ is the scalar  component of the chiral quark self energy.  As a corollary, the the two-body problem is solved almost completely 
once a nontrivial solution of the gap equation is found.

This remarkable fact facilitates the phenomenology of light pseudoscalar and is taken advantage of within the joint approach of the Dyson-Schwinger equation (DSE) 
and Bethe-Salpeter equation (BSE) in continuum Quantum Chromodynamics (QCD)~\cite{Cloet:2013jya,Krein:1990sf,Roberts:1994dr,Alkofer:2000wg,Fischer:2003rp,
Maris:2003vk,Roberts:2000aa,Bashir:2012fs}. That is because in both, the two-point and four-point Green functions, the axialvector Ward-Green-Takahashi identity is 
preserved by their simplest approximation, namely the rainbow-ladder (RL) truncation. There is no reason to expect {\em a priori\/} this truncation to be as successful 
in describing vector mesons whose solutions contain double the amount of covariants and whose higher masses sample the quark propagator in larger domain of the 
complex $p^2$ plane. Moreover, the axialvector Ward-Green- Takahashi  identity does not constrain the transverse components of the vector meson Bethe-Salpeter 
amplitude (BSA). Nonetheless, the simplest truncation carried out with the Maris-Tandy model~\cite{Maris:1997tm,Maris:1999nt}  for the quark-gluon interaction function 
works remarkably well for the lowest ground-state vector mesons, such as the $\rho$, $\omega$, $K^*$ and $\phi$ mesons. 

We here reassess the seminal work on vector mesons by Maris and Tandy~\cite{Maris:1999nt} within a modern understanding of the QCD 
interactions~\cite{Qin:2011dd,Qin:2013mta}. This ansatz produces an infrared behavior of the interaction, commonly described by a ``dressing function" $\mathcal{G}(k^2)$, 
that mirrors the decoupling solution found in DSE and lattice studies of the gluon propagator~\cite{Bowman:2005vx,Furui:2006ks,Cucchieri:2007zm,Cucchieri:2010xr,
Bogolubsky:2009dc,Sternbeck:2012mf,Oliveira:2012eh,Oliveira:2010xc,Boucaud:2011ug,Aguilar:2006gr,Aguilar:2008xm,Pennington:2011xs}. This solution is a bounded 
and regular function of spacelike momenta with a maximum value at $k^2 = 0$. Our aim is to compute the mass spectrum of the ground and radially excited states 
of the light, strange and charm vector mesons as well as of vector charmonia and bottomia~\cite{Serna:2016kdb,Serna:2017nlr,Bedolla:2015mpa,Bedolla:2016yxq,
Blank:2011ha,Krassnigg:2004if,Hilger:2015hka,Hilger:2015ora,Hilger:2016efh,Hilger:2017jti,Ding:2015rkn}  which were also the object of lattice-regularized QCD 
studies~\cite{Dudek:2010wm,Cheung:2016bym,Liu:2012ze};  in addition we compute their weak decay constants whose precise knowledge is important in hadronic 
observables measured by LHC$b$ and FAIR-GSI, for example.


\section{Bound States in the Vector Channel}

In analogy with previous work on pseudoscalar ground and excited states~\cite{Rojas:2014aka}, we employ the RL truncation in both, the quark's DSE and the vector meson's BSE, 
which is the leading term in a symmetry-preserving truncation scheme. The following two sections detail their respective kernels and lay out the setup for the numerical implementation
to compute vector meson properties.  

\subsection{Quark Gap Equation}
\begin{figure}[t!]
  \begin{center}
 	\includegraphics[width=0.5\textwidth]
 	{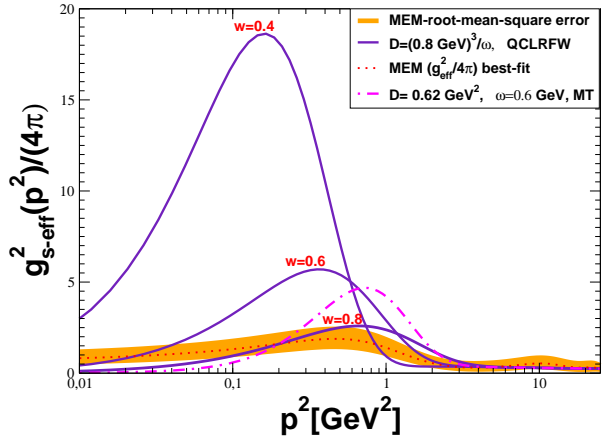}
\caption{(Color online) The solid indigo curves correspond to the effective coupling strength $\mathcal{G}$ in Ref.~\cite{Qin:2011dd} with $w D = (0.8~\text{GeV})^3$; 
the dashed-dotted magenta curve depicts the effective strength in the Maris-Tandy model~\cite{Maris:1999nt}; the dotted red curve corresponds  to the effective strength 
extracted  from lattice QCD data by using the maximum entropy method~(MEM); see Ref.~\cite{Rojas:2013tza} for  details.}
\label{fig:1}
 \end{center}
\end{figure}

The quark's gap equation is generally described by the DSE,\footnote{We employ throughout a Euclidean metric in our notation: 
$\{\gamma_\mu,\gamma_\nu\} = 2\delta_{\mu\nu}$; $\gamma_\mu^\dagger = \gamma_\mu$; $\gamma_5= 
  \gamma_4\gamma_1\gamma_2\gamma_3$, tr$[\gamma_4\gamma_\mu\gamma_\nu\gamma_\rho\gamma_\sigma]=-4\,
 \epsilon_{\mu\nu\rho\sigma}$; $\sigma_{\mu\nu}=(i/2)[\gamma_\mu,\gamma_\nu]$; 
$a\cdot b = \sum_{i=1}^4 a_i b_i$; and $P_\mu$ timelike $\Rightarrow$ $P^2<0$.}
\begin{align}
S^{-1}(p)  & =   \, Z_2 (i\, \gamma \cdot  p + m^{\mathrm{bm}}) \nonumber \\
                & +   \, Z_1\, g^2\!  \int^\Lambda_k \!  D^{\mu\nu} (q) \frac{\lambda^a}{2} \gamma_\mu\, S(k) \,\Gamma^a_\nu (k,p) \ ,
\label{QuarkDSE}
\end{align}
where $q=k-p$, $Z_{1,2}(\mu,\Lambda )$ are the vertex and quark wave-function renormalization constants, respectively, and $\int_k^\Lambda\equiv \int^\Lambda d^4k/(2\pi)^4$ 
represents throughout a Poincar\'e-invariant regularization of the integral with the regularization mass scale,  $\Lambda$. Radiative gluon corrections in the second term 
of Eq.~\eqref{QuarkDSE} add to the current-quark bare mass, $m^{\mathrm{bm}}(\Lambda)$, where the integral is over the dressed gluon propagator, $D_{\mu\nu}(q)$, and
the dressed quark-gluon vertex, $\Gamma^a_\nu (k,p)$; the SU(3) matrices, $\lambda^a$, are in the fundamental representation. The gluon propagator is purely 
transversal in Landau gauge,
\begin{equation}
   D^{ab}_{\mu\nu} (q) =  \delta^{ab} \left( g_{\mu\nu} - \frac{k_\mu k_\nu}{q^2} \right) \frac{\Delta ( q^2) }{q^2} \ ,
\end{equation}   
where $\Delta ( k^2)$ is the gluon-dressing function. In RL approximation, the quark-gluon vertex is simply given by its perturbative limit,
\begin{equation}
  \Gamma^a_\mu (k, p) =  \frac{\lambda^a}{2} \, Z_1 \gamma_\mu  \  ,
\end{equation}
and since we neglect the three-gluon vertex and work in the ``Abelian" version of QCD which enforces a Ward-Green-Takahashi identity~\cite{Maris:1997tm,Binosi:2014aea,Binosi:2016wcx}, 
$Z_1=Z_2$, we re-express the kernel of Eq.~\eqref{QuarkDSE},
\begin{equation}
  Z_1\, g^2  D_{\mu\nu} (q) \,  \Gamma_\mu (k, p) \ =  \  Z_2^2  \, \mathcal{G} (q^2) \, D_{\mu\nu}^\mathrm{free} (q) \,  \gamma_\mu \ ,
\label{DSEtrunc}  
\end{equation}
where we suppress color factors and $ D_{\mu\nu}^\mathrm{free}  (q) := \big ( g_{\mu\nu} -  q_\mu q_\nu/q^2 \big )/q^2$ is the free gluon propagator. An effective model 
coupling,  whose momentum-dependence  is congruent with DSE- and lattice-QCD results and yields successful explanations of numerous hadron 
observables~\cite{Qin:2011xq,Cloet:2013jya,Bashir:2012fs,Rojas:2014aka}, is given by the sum of two scale-distinct contributions:
\begin{equation}
\frac{\mathcal{G} (q^2)}{q^2} =   \frac{8\pi^2}{\omega^4}  D  e^{- q^2/\omega^2 } 
                             +  \frac{8\pi^2 \gamma_m\, \mathcal{F}(q^2) }{\ln \Big [\tau + \big(1 + q^2/\Lambda_\mathrm{QCD}^2 \big )^{\! 2} \Big ] } \, ,
\label{qinchang}
\end{equation}
The first term is an infrared-massive and finite {\em ansatz\/} for the interaction, where  $\gamma_m = 12/(33 - 2N_f )$, $N_f = 4$, $\Lambda_\mathrm{QCD} = 0.234$~GeV; $\tau = e^2 - 1$; 
and $\mathcal{F}(q^2) = [1 - \exp(-q^2/4m^2_t  ) ]/q^2$, $m_t = 0.5$~GeV.  The parameters $\omega$ and $D$ control the width and strength of the interaction, respectively.
At first sight they seem to be independent, yet a large collection of observables of ground-state vector and isospin-nonzero pseudoscalar mesons are practically insensitive  to variations 
of $\omega \in [0.4, 0.6]$~ GeV, as long as $D\omega$ = constant. The second term in Eq.~\eqref{qinchang} is a bounded, monotonically decreasing  continuation of the perturbative-QCD 
running coupling for all  spacelike values of $q^2$. The most important feature of this  {\em ansatz\/} is that it provides sufficient strength to realize DCSB and implements a confined-gluon 
interaction~\cite{Bashir:2012fs}. At $k^2 \gtrsim 2$~GeV$^2$, the perturbative component dominates the interaction. In Figure \ref{fig:1} we plot the interaction, $\mathcal{G} (q^2)$ in Eq.~\eqref{qinchang} 
for a typical value, $w D = (0.8~\text{GeV})^3$ and $\omega =0.4$~\cite{Rojas:2014aka}, employed in RL approximation as well as for other values of $\omega$ to illustrate the decrease of  
and shift towards  larger $k^2$ of its strength. For $\omega \simeq 0.6$, the functional form of the interaction is more akin to that employed in combination with a ghost-dressed Ball-Chiu 
vertex~\cite{Rojas:2013tza,El-Bennich:2013yna,Rojas:2014tya}.

The solutions for spacelike momenta, $p^2>0$, of the gap equation~\eqref{QuarkDSE} include a vector and a scalar piece,
\begin{equation}
   S_f^{-1}(p) = i\, \gamma \cdot  p\ A_f(p^2) +  \mathbf{1}_D\,  B_f(p^2)  \  ,
  \label{sigmaSV}
\end{equation}
for a given flavor, $f$, which requires a renormalization condition for the quark's wave function,
\begin{equation}
   \left. Z_f (p^2) = 1/A_f (p^2)  \right |_{p^2 = 4~\mathrm{GeV}^2} = 1 \ .
\label{EQ:Amu_ren}
\end{equation}
This imposed condition is supported by lattice-QCD simulations of the dressed-quark propagator. The mass function,  $M_f(p^2)=B_f(p^2\!, \mu^2)/A_f(p^2\!, \mu^2)$, is 
renormalization-point independent. In order to reproduce the quark-mass value in perturbative QCD, another renormalization condition is imposed,
\begin{equation}
\left.  S^{-1}_f(p) \right |_{p^2=\mu^2}  = \  i\  \gamma \cdot  p \ + \mathbf{1}_D\,  m_f(\mu )  \ ,
\label{massmu_ren}
\end{equation}
at a large spacelike renormalization point, $\mu^2\gg \Lambda_\mathrm{QCD}^2$, where $m_f(\mu )$ is the renormalized running quark mass:
\begin{equation}
\label{mzeta} 
   Z_m^f  (\mu,\Lambda )\, m_f(\mu)  =  m_f^{\rm bm} (\Lambda) \  . 
\end{equation}
Here,  $Z_m^f(\mu,\Lambda ) =  Z_4^f(\mu,\Lambda )/Z_2^f (\mu,\Lambda )$ is the flavor dependent mass-renormalization constant
and  $Z_4^f(\mu,\Lambda )$ is associated with the mass term in Lagrangian. In particular, $m_f(\mu )$  is nothing else but the dressed-quark mass 
function evaluated at one particular deep spacelike point, $p^2=\mu^2$, namely: $m_f(\mu)  = M_f(\mu )$.



\subsection{Vector Bound-State Equation}

The wave function of a bound state of a quark of flavor, $f$, and an antiquark of flavor, $\bar g$, in the $1^-$ channel is related to their BSA, $\Gamma_{V\mu}^{f\bar g}  (p,P)$,
which for a relative momentum, $p$, and total momentum, $P$, is obtained from the  homogeneous BSE,
\begin{equation}
\label{BSE} 
\hspace*{-2mm}
  \Gamma_\mu^V  (p,P)  =   \int^\Lambda_k \!  \mathcal{K} (p,k,P)\,  S_f (k_+)\, \Gamma_\mu^V (k,P)\, S_{\bar g} (k_-)  ,
\end{equation}
where $k_+ = k+\eta_+ P, k_- =k- \eta_- P; \eta_+ +\eta_- =1$.  We employ a ladder truncation of the BSE kernel consistent with that of the 
quark's DSE~\eqref{DSEtrunc},
\begin{equation}
\label{BSEkernel} 
    \mathcal{K} (p,k,P) =  - Z_2^2\, \mathcal{G} (q^2 )  \frac{\lambda^a}{2}  \gamma_\mu  \,  D^\mathrm{free}_{\mu\nu} (q) \, \frac{\lambda^a}{2}  \gamma_\nu \  ,
\end{equation}
which satisfies an axial-vector Ward-Green-Takahashi identity~\cite{Maris:1997hd} and consequently the pseudoscalar mesons are 
massless in the chiral limit. The BSE defines an eigenvalue problem with physical on-shell solutions for $P^2 = -M_{V_0}^2$ for the ground state and
for the radially excited states, $P^2 = - M^2_{V_n}$, $M_{V_{n+1}}^2 > M_{V_n}^2$, $n=1,2,3 ...$.

As we are interested in radially excited $J^P = 1^-$ states, the question arises whether they can be described by the interaction in Eq.~\eqref{qinchang} 
with  exactly the same parameter set as for ground states, whether the parameters have to be adjusted or whether the truncation fails to achieve
at least a reasonable description of their mass spectrum. The masses and weak decay constants of excited states are very sensitive to the strength
and width of the long-range term in Eq.~\eqref{qinchang}, which provides more support at large inter-quark separation than, e.g., the Maris-Tandy 
model~\cite{Maris:1999nt}. In here, we  extend the studies of Refs.~\cite{Qin:2011dd,Rojas:2014aka} to the excited states of vector mesons
with an interaction that differs from those employed in Ref.~\cite{Hilger:2017jti}. We also refer to the discussion in Ref.~\cite{El-Bennich:2016qmb},
where it is pointed out that beyond-RL contributions are important in heavy-light mesons due to the strikingly different impact of the quark-gluon 
vertex dressing for a light and a heavy quark. The effects of DSCB and the importance of other quark-gluon  tensor structures are increasingly more 
important for lighter quarks~\cite{Bashir:2012fs,Binosi:2016wcx}. Thus, one does not expect the RL truncation to accurately describe either the ground 
nor the excited states of charm and beauty mesons. On the other hand,  the RL approximation describes very well equal-mass bound states, such as 
quarkonia~\cite{Serna:2017nlr,Bedolla:2015mpa,Bedolla:2016yxq,Krassnigg:2004if,Blank:2011ha,Hilger:2015hka,Hilger:2015ora,Hilger:2016efh,Hilger:2017jti}.

The normalization condition for the Bethe-Salpeter amplitude is,
\begin{align}
   2P_{\mu} \ = \ & \frac{\partial}{\partial P_{\mu}}\frac{N_{c}}{3}  \int^{\Lambda}_k   \mathrm{Tr_D}  \Big [  \bar \Gamma_\nu^V  (k, - K)S_f (k_+)    \notag\\
                    \times \ & \Gamma_\nu^V (k, K) S_{\bar g}  (k_-)  \Big ]_{K=P}^{P^2=-M^2_V},
\label{norm}
\end{align}
The charge-conjugated BSA is defined as $\bar \Gamma (k,-P) := C\, \Gamma^T (-k,-P) C^T$, where $C$ is the charge conjugation operator. 

Finally, the weak decay constant for $1^{-}$ meson is defined as,
\begin{equation}
   f_V  M_V  \epsilon^\lambda_\mu (P)  = \langle 0 | \bar q^{\bar g} \gamma_\mu q^f |  V  (P,\lambda) \rangle  \ ,
\label{decaydef}
\end{equation}
where $\epsilon^\lambda_\mu (P)$ is the meson's polarization vector satisfying  $\epsilon^\lambda_\mu \cdot P = 0$ and
normalized such that ${\epsilon^{\lambda}_\mu}^* \!\cdot \epsilon^\lambda_\mu = 3$; Eq.~\eqref{decaydef} can be expressed as,
\begin{equation}
\hspace*{-2.8mm}
 f_V  M_V  = \frac{Z_2N_c}{3} \!\!\int^{\Lambda}_k  \!\! \mathrm{Tr}_D \big [  \gamma_\mu  S_f (k_+) \Gamma_{V\mu}^{f\bar g} (k, K) S_{\bar g}  (k_-) \big ].
\label{vectordecay}
\end{equation}


\section{Numerical Implementation \label{numerical}}

\subsection{Quark Propagators on the Complex Plane \label{complexplane} }

In solving the BSE~\eqref{BSE}, the quark propagators with momentum $(k\pm P)^2 = k^2 + 2i \eta_\pm |k| M_V - \eta_\pm^2 M_V^2$,  where $k$ is collinear with $P = (\vec 0,i M_V )$ in the meson's 
rest frame, must necessarily be treated in the complex plane~\cite{Rojas:2014aka,Hilger:2017jti}. Complex-conjugate pole positions of the propagators depend on the analytical form of the 
interaction and  can be represented by analytical expressions based on a complex-conjugate pole model~\cite{Bhagwat:2002tx}:
\begin{equation}
    S(p) = \sum_i^n \left [ \frac{z_i}{i\gamma\cdot  p + m_i }  + \frac{z_i^*}{i\gamma \cdot  p + m_i^*}\right ]  \ ; \ m_i, z_i \in \mathbb{C} \ .
    \label{pole}
\end{equation}
The propagator in Eq.~\eqref{pole} is pole-less on the real timelike axis and therefore has no K\" all\' en-Lehmann representation, which is a sufficient condition to implement 
confinement~\cite{Krein:1990sf,Cloet:2013jya}. The numerical DSE solutions we obtain on the complex plane~\cite{Rojas:2014aka,El-Bennich:2016qmb} can be fitted with $n=3$ 
complex-conjugate poles. We solve the BSE~\eqref{BSE} both ways, employing full numerical DSE solutions for the quark in the complex plane and the pole model in Eq.~\eqref{pole}
and find agreement at the one-percent level for the vector meson masses and decay constants. 

With respect to the current-quark masses given by,
\begin{equation}
   Z_4^f (\mu, \Lambda ) \, m_f (\mu) =   Z_2^f (\mu, \Lambda ) \, m_f^\mathrm{mb} (\Lambda) \ ,
\end{equation}
where $ Z_4^f (\mu, \Lambda ) $ is  associated with the mass term in the QCD Lagrangian, $m_u=m_d(\mu) , m_s(\mu) $ and $m_c(\mu) $  are fixed in Eq.~\eqref{QuarkDSE} 
by requiring  that the pion and kaon BSEs produce $m_\pi = 0.138$~GeV and $m_K = 0.493$~GeV, respectively. This, in turn, yields $m_{u,d}(\mu) = 3.4~\mathrm{MeV}$, 
$m_s(\mu) = 82~\mathrm{MeV}$,  $m_c(\mu) =0.828 $~GeV  and $m_b(\mu) = 3.86$~GeV for $\mu =19$~GeV.

\subsection{Solving the Bethe-Salpeter Equation}

The general Poincar\'e-invariant form of the solutions of Eq.~(\ref{BSE}) in the vector meson channel and for the eigenvalue trajectory, $P^2= -M_{V_n}^2$, in a orthogonal base with 
respect to the Dirac trace is given by:
\begin{align}    
   \Gamma^{V_n}_\mu (q;P)=\sum^{8}_{\alpha=1} T^{\alpha}_{\mu}(q, P)\, \mathcal{F}^n_{\alpha}(q^{2},q\cdot P;P^{2}) \, ,
\label{diracbase}                            
\end{align}
with the dimensionless orthogonal Dirac basis~\cite{Maris:1999nt},
\begin{eqnarray}
T^{1}_{\mu}(q,P)  &=& \gamma^{T}_{\mu}\, , \\
T^{2}_{\mu}(q,P)  &=& \frac{6}{q^{2}\sqrt{5}} \Big [q^{T}_{\mu}(\gamma^T\! \cdot q)-\frac{1}{3}\gamma^{T}_{\mu}(q^{T})^{2} \Big ] \, ,  \\
T^{3}_{\mu}(q,P)  &=& \frac{2}{qP} \, \Big [q^{T}_{\mu}(\gamma\cdot P) \Big ]\,  , \\
T^{4}_{\mu}(q,P)  &=& \frac{i\sqrt{2}}{qP}  \Big [\gamma^{T}_{\mu}(\gamma\cdot P)(\gamma^T\! \cdot q)+q^{T}_{\mu}(\gamma\cdot P)  \Big ] \, ,  \hspace*{5mm} \\
T^{5}_{\mu}(q,P)  &=& \frac{2}{q}\, q^{T}_{\mu}\, , \\
T^{6}_{\mu}(q,P) &=& \frac{i}{q\sqrt{2}} \Big [ \gamma^{T}_{\mu}(\gamma^T\! \cdot q)-(\gamma^T \! \cdot q)\gamma^{T}_{\mu} \Big ]  \,  , \\
T^{7}_{\mu}(q,P)  &=& \frac{i\sqrt{3}}{P\sqrt{5}} \left ( 1-\cos^2{\theta} \right )  \Big [  \gamma^{T}_{\mu}(\gamma\cdot P)  \\
                            &- & (\gamma\cdot P)\gamma^{T}_{\mu}  \Big ]-\frac{1}{\sqrt{2}}\, T^{8}_{\mu}(q,P)\,  ,  \\
T^{8}_{\mu}(q,P)  &=& \frac{i \, 2\sqrt{6}}{q^{2}P\sqrt{5}}\ q^{T}_{\mu} \, \gamma^T \!\!\cdot q  \ \gamma\cdot P \,  .
\end{eqnarray}
The $C$-parity properties of this basis are elucidated elsewhere~\cite{Maris:1999nt} and the transverse projection, $V^{T}$, is defined by,
\begin{equation}
   V^{T}_{\mu}= V_{\mu}-\frac{P_{\mu}(P\cdot V)}{P^{2}},
\end{equation}
with $q\cdot P=qP\cos{\theta}$.
These covariants satisfy the orthonormality condition,
\begin{equation}
   \tfrac{1}{12}\,  \text{Tr}_{\text{D}} \big [ T^{\alpha}_{\mu}(q,P)\, T^{\beta}_{\mu}(q, P) \big ]   = f_{\alpha}(\cos{\theta})\delta^{\alpha\beta} \, ,
\label{orthorel}
\end{equation}
where the functions $f_{\alpha}(z)$ are given by  $f_{1}(z)=1$, $f_{\alpha}(z)= \frac{4}{3}(1-z^2)$ with  $\alpha=3,4,5,6$  and  $f_{\alpha}(z)= \frac{8}{5}(1-z^2)^2$ for  $\alpha=2,7,8$.
The normalization constants $f_\alpha$ satisfy: 
\begin{equation}
    \int_0^\pi \! d\theta\, \sin^2\! \theta \, f_{\alpha}(\cos{\theta})=\frac{\pi}{2} \ .
\end{equation}
Making use of the covariant decomposition in Eq.~\eqref{diracbase} and the orthogonality relations~\eqref{orthorel}, the homogeneous BSE~\eqref{BSE} with the kernel~\eqref{BSEkernel} 
can be recast in a set of eight  coupled-integral equations,
\begin{align}
  &\mathcal{F}^n_{\alpha}(p^{2},p\cdot P, P^{2})f_{\alpha}(z)= \nonumber  \\ 
  &-\tfrac{4}{3}  \int^\Lambda_k  \! \mathcal{G}\left( q^{2}\right)D^{\mathrm{free}}_{\mu\nu}(q)\, \mathcal{F}^n_{\beta} (k^{2},k\cdot P;P^{2})  \nonumber \\
  &  \times\,  \tfrac{1}{12}\,  \mathrm{Tr}_{\text{D}}\big [ T^{\alpha}_{\rho}(p;P) \gamma_{\mu}S_f (k_+)  T^{\beta}_{\rho}(k;P)S_{\bar g }(k_-)  \gamma_{\nu} \big ]  \, . 
   \label{coupledeq}
\end{align}
This equation can be posed as an eigenvalue problem for a set of eigenvectors $\bm{\mathcal{F}^n} := \{\mathcal{F}_\alpha^n; \alpha =1,...,8 \} $:
\begin{equation}
 \lambda_n (P^2) \bm{\mathcal{F}^n} = \mathcal{K} (p,k,P)\, \bm{\mathcal{F}^n}  \ .
 \label{eigenvalue}
\end{equation}
For every solution eigenvector, $\bm{\mathcal{F}^n}$, there exists a mass, $M_{V_n}$, such that $\lambda_n(-M^2_{V_n})=1$~\cite{Rojas:2014aka,Qin:2011xq,Holl:2004fr,
El-Bennich:2015kja,Segovia:2015hra,Afonin:2014nya}. The set of masses, $M_{V_n}$, represents the radially excited meson spectrum of quark-antiquark bound states with $J^P= 1^{-}$. 
In order to improve a faster convergence in solving the coupled equations~\eqref{coupledeq},  we expand the eigenfunction into Chebyshev polynomials,
\begin{equation}
  \mathcal{F}_{\alpha}^n  (k^{2},k\cdot P;P^{2})=\sum^{\infty}_{m=0} \mathcal{F}_{\alpha m}^n  (k^{2};P^{2})U_{m}(z_k),
\end{equation}
where the  $U_{m}(z)$ are  Chebyshev polynomials of second kind and the angles, $z_k = P \cdot k/(\surd{P^2}\surd{k^2})$ and $z_p = P \cdot p/(\surd{P^2}\surd{p^2})$,
and momenta, $k$ and $p$, are discretized~\cite{Rojas:2014aka}. We employ three Chebyshev polynomials for the ground and five for excited  states. 
We solve the eigenvalue problem posed in Eq.~(\ref{eigenvalue}) by means of the implicitly restarted Arnoldi method, as implemented in the \texttt{ARPACK} library~\cite{Arpack}
which computes the eigenvalue spectrum for a given $N\times N$ matrix. A practical implementation requires a mapping of the BSE kernel $\mathcal{K}_{\alpha\beta} (p,k,P)$ 
onto such a square matrix and is described in detail in Ref.~\cite{Rojas:2014aka}. 
   
%

We obtain the eigenvalue spectrum,  $\lambda_n (P^2)$, of the kernel in Eq.~\eqref{coupledeq} and the associated eigenvectors, $\bm{\mathcal{F}^n}$ of the
vector meson's BSAs where the root, $M_{V_n}$, of the equation $\lambda_n (P^2=-M_{V_n}^2)-1=0$ is found by employing the  {\em Numerical Recipe\/}~\cite{Press:1992zz}
subroutines \texttt{zbrent} and \texttt{rtsec}. We verify the \texttt{ARPACK} solutions with the commonly used iterative procedure and find excellent agreement of the 
order $10^{-16}$.


\section{Discussion of results \label{discussion}}

We summarize our results for the mass spectrum and weak decay constants of the flavor-singlet and  light-flavored vector mesons in 
Tables~\ref{table1}  and \ref{table2}, where the DSE and BSE are solved for two interaction, $\mathcal{G}(q^2)$, parameter sets in 
Eq.~\eqref{qinchang}, namely $\omega = 0.4$~GeV and $\omega = 0.6$~GeV and the fixed value $\omega D =(0.8~\mathrm{GeV})^3$;
see also discussion in Ref.~\cite{Qin:2011xq}. In Table~\ref{table1}, we list the $1^-$ masses for the ground state and first radial excitation
following the Particle Data Group (PDG)  conventions~\cite{Olive:2016xmw}, whereas in Table~\ref{table2} this is done for the weak
decay constants. 

A direct comparison of the mass and decay constant entries in both model-interaction columns reveals that the values obtained with $\omega = 0.4$~GeV 
are in much better agreement with experimental values of the $1^-$ ground states, namely in case of the $\rho$, $K^*(892)$, $\phi(1020)$ and  $J/\psi$, 
whose  $\omega$ dependence in the range $\omega \in [0.4,0.6]$~GeV is rather weak. On the other hand, for $\omega = 0.6$~GeV  the masses obtained
for the radially excited states, $\phi(1680)$ and  $\psi(2S)$, are only marginally better. It turns out that in the RL approximation, a realistic description
of the radially excited vector meson masses is not possible with $\omega D =(0.8~\mathrm{GeV})^3$.

\begin{table}[t!]
\centering
\def\arraystretch{1.5}
\begin{tabular}{|c|c|c|c|} \hline
$J^P = 1^-$        &$M_{V_n}^{\omega=0.4}$& $M_{V_n}^{\omega=0.6}$ & $M_{V_n}^{\text{exp}}$~\cite{Olive:2016xmw} \\ \hline
$\rho^0(770)$ & 0.742 & 0.695  & 0.775   \\ 
$\rho^0(1450)$& 0.942 & 0.927  & 1.465     \\ 
$K^{*}(892)$  & 0.951 &  0.914  & 0.896    \\ 
$K^{*}(1410)$ & 1.217& 1.206  & 1.414   \\ 
$\phi(1020)$  & 1.087 & 1.055  & 1.019  \\ 
$\phi(1680)$  & 1.295 & 1.376  & 1.659  \\ 
$J/\psi$      & 3.114 & 3.065  & 3.097  \\ 
$\psi(2S)$    & 3.393 & 3.507  & 3.689  \\ 
$\Upsilon(1S)$    & 9.634 & 9.552  & 9.460   \\ 
$\Upsilon(2S)$& 9.945 & 9.848  &10.023   \\ \hline
\end{tabular}
\caption{ Mass spectrum~[in GeV] of flavor singlet and non-singlet mesons in the $J^{P}=1^-$ channel. The model parameter $\omega$ refers to the interaction {\em ansatz} 
in Eq.~\eqref{qinchang}, and we exemplify the spectrum for the values $\omega = 0.4$~GeV and $\omega = 0.6$~GeV with $\omega D =(0.8~\mathrm{GeV})^3$ fixed.
We consider the ground state and first radial resonance, $n=0,1$, and compare with experimental values of the PDG~\cite{Olive:2016xmw} whose conventions
we use in the last column.}
\label{table1}
\end{table}


\begin{table}[b!]
\centering
\def\arraystretch{1.5}
\begin{tabular}{|c|c|c|c|} \hline
$J^P = 1^-$      &  $f_{V_n}^{\omega=0.4}$  &   $f_{V_n}^{\omega=0.6}$ & $f_{V_n}^{\text{exp.}}$        \\ \hline
$\rho^0(770)$ & 0.231 & 0.242 & 0.221  \\ 
$\rho^0(1450)$&  ---  & --- &       \\ 
$K^{*}(892)$  & 0.287 & 0.304 & 0.217   \\ 
$K^{*}(1410)$ & 0.195 & 0.127 &         \\ 
$\phi(1020)$  & 0.299 & 0.305 & 0.322  \\ 
$\phi(1680)$  & 0.102 & 0.061 &         \\ 
$J/\psi$      & 0.433 & 0.463 & 0.416  \\ 
$\psi(2S)$    & 0.208 & 0.230 & 0.295  \\ 
$\Upsilon(1S)$&  --- & --- & 0.715        \\ 
$\Upsilon(2S)$&  --- & --- & 0.497        \\ \hline
\end{tabular}
\caption{ Weak decay constants [in GeV] of flavor singlet and non-singlet $J^{P}=1^-$ mesons; see Table~\ref{table1} for explanations. Reference values 
for $f_{V_n}$ are listed in the last column when available~\cite{Olive:2016xmw}. The long dash stands for numerically unstable results; i.e. the
integral expression~\eqref{vectordecay} does not stabilize with increasing numbers of Chebyshev moments. In the last column, experimental decay 
constants are extracted from the PDG values~\cite{Olive:2016xmw} using the formulae in Appendix~\ref{apendice}. }
\label{table2}
\end{table}


We thus choose $\omega D = (1.1~\mathrm{GeV})^3$ and $\omega = 0.6$~GeV for which the numerical mass and decay constant values of 
the radially excited states are presented in Table~\ref{table3} and compare well with experimental values, yet the ground
states are no longer insensitive to $\omega$ variations for $\omega D = (1.1~\mathrm{GeV})^3$~\cite{Qin:2011xq}. In order to maintain
$m_\pi = 0.138$~GeV,  $\omega$ must increase beyond our reference value, $\omega = 0.6$~GeV, for the excited spectrum. These results 
confirm an analogous trend observed for pseudoscalar mesons~\cite{Rojas:2014aka}. Nonetheless, the ground states are noticeably less 
dependent on the $\omega D$ values than the radial excitations where large  mass differences are observed between both parameter sets. 
This agrees with the observations made in Refs.~\cite{Qin:2011xq,Rojas:2014aka} and extends them to the strange and charm vector mesons:  
the quantity $r_\omega := 1/\omega$  is a length scale that measures the range of the interaction's infrared component in Eq.~\eqref{qinchang}. 
The radially excited states were shown to be more sensitive to long-range characteristics of $\mathcal{G}(q^2)$ than the ground states and we 
confirm  that the mass of the radially excited states is lowered when $r_\omega$ decreases except in case of the $\phi(1680)$ and  $\psi(2S)$ mesons. 

The masses of the ground and the radial excitation states of the vector mesons we find correspond to the first and third eigenvalues (from highest to lowest),
respectively. This is because the second eigenvalue does not correspond to $1^{--}$ states since the even Chebyshev moments are strongly suppressed.  
The exceptions are the $\rho(1450)$ using $D\omega = (1.1~\text{GeV})^3$  and the $K^*(1410)$ with $D\omega = (0.8~\text{GeV})^3$  and $\omega =0.6 $ 
where the radial  excitation {\em does\/} correspond to the second eigenvalue. (NB: the radial excitations have identical quantum numbers as the ground 
state; therefore, the odd Chebyshev polynomials must be suppressed as it occurs for the ground states).


\begin{table}[t]
\centering
\def\arraystretch{1.5}
\begin{tabular}{|c|c|c|c|c|} \hline
$J^P = 1^-$  & $M_{V_n}$ & $f_{V_n}$ &  $M_{V_n}^{\text{exp}}$&  $f_{V_n}^{\text{exp.}}$ \\ \hline
$\rho(1410)$  & 1.284    & 0.150   & 1.465 &  ---        \\
$\phi(1680)$&  1.650   &  0.138   & 1.659 &  ---         \\
$\psi(2s)$  & 3.760    & 0.176    & 3.689 & 0.295   \\ 
$\Upsilon(2s)$&10.140   &  0.564  & 10.023 &   0.497    \\
\hline
\end{tabular}
\caption{ Mass spectrum and weak decay constants for the first radially excited flavorless $J^{P}=1^-$ states following PDG
conventions. All values are in GeV and obtained with the interaction in Eq.~\eqref{qinchang} and the  parameter values  $\omega = 0.6$~GeV 
and  $\omega D =(1.1~\text{GeV})^3$. The long dash denotes numerically unstable results. In the first and fifth columns,  experimental 
masses~\cite{Olive:2016xmw} and  reference values for the  decay constants $f_{V_n}$ are given when available. 
Experimental values for $f_{V_n}$ are extracted from the PDG~\cite{Olive:2016xmw} using the formulae in Appendix~\ref{apendice}. }
\label{table3}
\end{table}

In summary, we do not find a parameter set that describes equally well the entire  mass spectrum of ground and excited states, which
demonstrates the insufficiency of this truncation and confirms our finding in the pseudoscalar channel~\cite{Rojas:2014aka}.


\section{Conclusion}

We computed the BSAs for the ground and first excited states of the flavor-singlet and light-flavored vector  mesons with an interaction
{\em ansatz} that is massive and finite in the infrared and massless in the ultraviolet domain. This interaction is qualitatively in accordance 
with the so-called {\em decoupling solutions\/} of the gluon's dressing function and thus supersedes the Maris-Tandy model~\cite{Maris:1999nt}
that vanishes at small momentum squared. In conjunction with the RL truncation, the latter proved to be a successful interaction model for the flavorless 
light pseudoscalar and vector mesons as well as quarkonia.

Motivated by the successful application of this interaction to the mass spectrum of light vector mesons as well as  some of their excited states in Ref.~\cite{Qin:2011xq}, 
we extend this study to the strange and charm sectors and obtain the masses of ground and radially excited states as presented in  Table~\ref{table1}
and also compute their weak decay constants. The numerical values obtained are in good agreement with experimental data in case of ground states, 
but the same parametrization yields values that compare poorly with experiment for the excited states.  We thus confirm our earlier observation that no 
single parametrization  of Eq~\eqref{qinchang} is suitable to reproduce the mass spectrum of both, the ground and excited states in RL truncation. 
Although not explicitly detailed here, this approximation also fails to produce the correct masses for the $D_{(s)}$ and $B_{(s)}$ vector mesons and
the discrepancy is even more pronounced than in the case of charmed pseudoscalar mesons~\cite{Rojas:2014aka}. Reasons for this were 
put forward, e.g., in Ref.~\cite{El-Bennich:2016qmb}. 

It has therefore  become strikingly clear that a unified description of flavored pseudoscalar and vector mesons, quarkonia and their radial excitations can only be
achieved within a treatment of the BSE beyond the leading truncation.

\acknowledgements 

The work of F.~Mojica and C.~E.~Vera was supported by {\em Comit\'e Central de Investigaciones, Universidad del Tolima\/}, under project no.~60220516.
E.~Rojas acknowledges financial support from {\em Patrimonio Aut\'onomo Fondo Nacional de Financiamiento para la Ciencia, la Tecnolog\'ia y la 
Innovaci\'on, Francisco Jos\'e de Caldas\/} and {\em Sostenibilidad-UDEA}. B. El-Bennich is supported by the S\~ao Paulo Research Foundation 
(FAPESP), grant no.~2016/03154-7, and by {\em Conselho Nacional de Desenvolvimento Cient\'ifico e Tecnol\'gico} (CNPq), grant no.~458371/2014-9.

\appendix

\section{Extraction of the decay constants}

\label{apendice}
 Following Ref.~\cite{Maris:1999nt,Choi:2015ywa}, 
we can extract the  decay constant of the  vector mesons from  the experimental value~\cite{Olive:2016xmw} for  the partial width of the $\rho, V \longrightarrow  e^{+}e^{-}$ decay,
where $V$ denotes the $\phi$ and heavy-flavored  mesons.   
\begin{align}
f_{\rho}^2=& \frac{3 m_{\rho} }{2\pi \alpha^2}\Gamma_{\rho\rightarrow e^{+}e^{-}},\notag\\
f_{V}^2=&  \frac{3 m_{V} }{4\pi \alpha^2 Q^2}\Gamma_{V\rightarrow e^{+}e^{-}}. 
\end{align}
In this expression, $Q$ is the charge of the quarks  in the meson and  $\alpha$ is the fine structure constant.

\end{document}